%% file: TBH3.tex
\documentclass[floatfix,twocolumn,showpacs,nofootinbib,prd,aps,tightenlines]{revtex4-1}
\usepackage{graphicx}
\usepackage{psfrag}
\usepackage{dcolumn}
\usepackage{bm}
\usepackage{amsmath}
\usepackage{bm}
\usepackage{appendix}
\usepackage[caption=false]{subfig}

\newcommand{\bea}{\begin{eqnarray}}
\newcommand{\eea}{\end{eqnarray}}
\newcommand{\beq}{\begin{equation}}
\newcommand{\eeq}{\end{equation}}

\usepackage{color}

\begin{document}

\title{Close Encounter of Three Black Holes III}

\author{Giuseppe Ficarra}
\affiliation{Center for Computational Relativity and Gravitation,
School of Mathematical Sciences,
Rochester Institute of Technology, 85 Lomb Memorial Drive, Rochester,
New York 14623, USA}
\author{Carlos O. Lousto}
\affiliation{Center for Computational Relativity and Gravitation,
School of Mathematical Sciences,
Rochester Institute of Technology, 85 Lomb Memorial Drive, Rochester,
New York 14623, USA}

\date{\today}

\begin{abstract}
We revisit the three black hole scenario with numerical relativity techniques
to study hierarchical configurations where the inner binary contains highly spinning black holes.
We find that the merger time of the binary gets a delay
(with a number of orbits to merger increase), depending strongly with the distance
to the orbiting third hole $D$ as $\sim1/D^{1.6\pm0.1}$. Notably,
a different dependence from what we had found in the nonspinning case, $\sim1/D^{2.5}$.
We interpret this effect as mostly due to a spin-orbit
coupling between the third hole
and the closest member of the binary in the successive approaches.
This lead us next to study scattering configurations of the third
hole with the binary
in order to evaluate the extent of this ``sudden'' interactions,
finding also a correlation of the delay in merger times with the
closest distance, even for the nonspinning cases.
We then explore the mass ratio dependence of the triple system by modeling binaries
orbiting a larger black hole bearing masses ratios 8:1:1 and 18:1:1 in co-orbiting
or counter-orbiting configurations,
finding merger times increasing with increasing mass ratios and for the counter-orbiting cases.
\end{abstract}

\pacs{04.25.dg, 04.25.Nx, 04.30.Db, 04.70.Bw}\maketitle

\section{Introduction}\label{sec:Intro}
There is a renewed interest in the study of multi black hole interactions,
in particular the effect a third black hole may have on a close binary.
Three body encounters and accretion effects
(see, e.g.,\cite{Schnittman:2015eaa, Nixon:2012zb, Antonini:2012ad, Samsing:2013kua,Bonetti:2017dan}) 
can lead to highly eccentric binaries, with residual eccentricity surviving down to near
merger. These eccentric binaries may have very interesting gravitational waves
signals that cannot 
be adequately modeled using quasicircular approximations \cite{Gayathri:2020coq}. 
This subject has been the focus of much interest lately
\cite{ShapiroKey:2010cnz,DOrazio:2018jnv,Hoang:2019kye}, 
but its detailed modeling is largely incomplete.
Reliable evolutions that include those effects between LISA and third generation detectors
sensitivity bands can also be used to exploit multiband observational opportunities
\cite{Sesana:2016ljz,Vitale:2016rfr,Barausse:2016eii,Bonetti:2017lnj,Bonetti:2018tpf}.

This paper is the third installment of the three black hole
interactions study using full Numerical General Relativity techniques.
In a first paper \cite{Lousto:2007rj} we have performed a set of proofs of principle
evolutions to show that effectively our moving punctures approach
\cite{Campanelli:2005dd} can evolve accurately multi black hole systems beyond binaries.
Next we have revisited the three black hole (3BH)
scenario in \cite{Ficarra:2023zjc} to study the influence of a third hole
on eccentricity generation during evolution and merger times of an inner nonspinning binary.
While the eccentricity evolution seemed mostly unaffected,
presenting a steady decay, the merger times could be modeled with a distance $D$,
to the third hole as a dependence $\approx1/D^{2.5}$.

In this paper we complete our intended trilogy of 3BH studies,
by revisiting the scenario of the third hole
interacting with an inner binary, this time with highly spinning black holes to
see if the previous results regarding merger time and eccentricity evolutions still stands.
We will also consider here the effects of 
a scattering third hole instead of in bound orbits and
finally the effects of a much more massive central third hole with
a binary orbiting around it.

\section{Full Numerical Techniques}\label{sec:FN}

In order to perform the full numerical 
simulations of three black holes we employ the LazEv code\cite{Zlochower:2005bj}
including 8th order spatial finite differences \cite{Lousto:2007rj}, 4th order
Runge-Kutta time integration, and a Courant factor $(dt/dx=1/4)$.

Regularly to compute the numerical initial data for binary black holes,
we use the puncture
approach~\cite{Brandt97b} along with the {\sc  TwoPunctures}
~\cite{Ansorg:2004ds} code, that we will use in a hybrid approach in this paper
as described in next Sec.~\ref{sec:ID}.
We use the {\sc AHFinderDirect}~\cite{Thornburg2003:AH-finding} code to locate
apparent horizons and compute horizon masses from its area $A_H$.
We also measure the magnitude of the horizon spin 
$S_H$, using the ``isolated horizon'' algorithm 
as  implemented in Ref.~\cite{Campanelli:2006fy}.

During evolution of the holes we use the
{\sc Carpet}~\cite{Schnetter-etal-03b} mesh refinement driver
which provides a ``moving boxes'' style of mesh refinement. In this
approach, refined grids of fixed size are arranged about the
coordinate centers of the holes.  The code then moves these fine
grids about the computational domain by following the trajectories of
the black holes.

The grid structure of our mesh refinements have a size of the largest
box for typical simulations of $\pm400M$.  The number of points between 0
and 400 on the coarsest grid is XXX in nXXX (i.e. n100 has 100
points).  So, the grid spacing on the coarsest level is 400/XXX.  The
resolution in the wavezone is $100M/$XXX (i.e. n100 has $M/1.00$, n120
has $M/1.2$ and n144 has $M/1.44$) and the rest of the levels is
adjusted globally. For comparable masses and non-spinning black holes, the grid around one of the black holes
($m_1$) is fixed at $\pm0.8M$ in size and is the 9th refinement level.
Therefore the grid spacing at this highest refinement level is 400/XXX/$2^8$. When considering small mass ratio binaries, we progressively add internal grid refinement levels \cite{Lousto:2020tnb}. In the case of spinning black holes we add an additional refinement level inside the black hole apparent horizon and we increase the global resolution to n120.

The extraction of gravitational radiation from the numerical
relativity simulations is performed using the formulas (22) and (23)
from \cite{Campanelli:1998jv} for the energy and linear momentum
radiated, respectively, and the formulas in \cite{Lousto:2007mh}
for angular momentum radiated, all in terms of the extracted Weyl scalar $\Psi_4$
at the observer location $R_{obs}=113M$.


\section{Approximate initial data}\label{sec:ID}

In Ref.~\cite{Ficarra:2023zjc} we have performed three black holes
evolution studies from full approximate initial data based on work in
\cite{Lousto:2007rj}, which extended first order expansion to include terms
of the sort $\vec{S}_i\times\vec{P}_i$ representing interactions of spin with
linear momentum, to higher order on those intrinsic
parameters of the holes, $\vec{S}_i$ and $\vec{P}_i$ as well as the inverse
of the distance of the holes.

Since in the Bowen-York approach to initial data \cite{Bowen:1980yu},
with the ansatz of conformal flatness of the three-metric and transverse
traceless of the extrinsic curvature, the momentum constraint is solved exactly,
we have to look for a perturbative solution of
the Hamiltonian constraint equation, reduced to the partial
differential equation for the conformal factor $\phi$
\begin{equation}
\Delta \phi = -\frac{1}{8}\phi^{-7}\hat{A}^{ij}\hat{A}_{ij},
\end{equation}
with the Bowen-York conformal extrinsic curvature solution to the momentum constraint
\begin{eqnarray}
  \hat{A}^{ij} = \sum_a^{N_{BHs}}\left(\frac{3}{2r_a^2}\left[2P_a^{(i}n_a^{j)} + (n_a^in_a^j-\eta^{ij})P_{ak}n_a^k\right]\right.\nonumber\\
  \left.+ \frac{6}{r_a^3} n_a^{(i}\epsilon^{j)kl}J_{ak}n_{al}\right),
\end{eqnarray}
where we label the momentum and the spin of the holes as 
$\textbf{P}_i$ and $\textbf{J}_i$, following the notation of \cite{Lousto:2007rj}.

For the purpose to find an approximate solution to the Hamiltonian constraint
we start from the analytical solution at order 0th given by
\begin{equation}
\phi_0 = 1 + \sum_a^{N_{BHs}}\frac{m_a}{2 r_a},
\end{equation}
which solves
\begin{equation}
\Delta \phi_0 = 0.
\end{equation}

To find the first perturbative order $u$ of the solution defined as
\begin{equation}
\phi = \phi_0 + u\ ,
\end{equation}
we consider the equation for $u_1$
\begin{equation}\label{Perturbation_1}
\Delta u_1 = -\frac{1}{8}\phi_0^{-7} \hat{A}^{ij}\hat{A}_{ij}.
\end{equation}

Then to solve for the second order piece $u_2$, we look at the
perturbation equation for a single black hole
we have \cite{Ficarra:2023zjc},
\begin{equation}
    \Delta u_2 = \frac{7}{8} \phi_0^{-8}u_1 \hat{A}^{ij}\hat{A}_{ij}.
\end{equation}

In Ref.~\cite{Ficarra:2023zjc} we have superposed those perturbative
solutions to the Hamiltonian constraint to study the evolution of
three equal mass, nonspinning, black hole systems in a hierarchical
configuration consisting of a close binary with a third hole in coplanar as well
as in polar and other precessing cases. Here we would like to revisit
the problem in the case the black holes in the inner binary carry a
significant spin. 

\begin{figure}
\includegraphics[angle=0,width=.9\columnwidth]{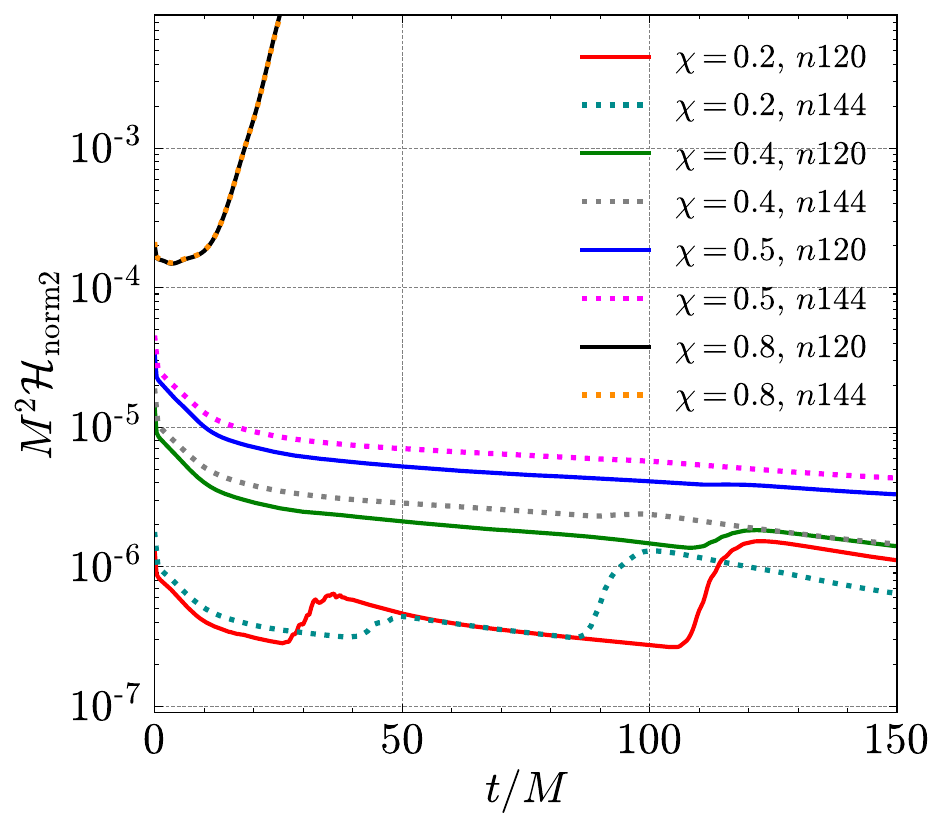}
  \caption{Evolution of the L2-norm of the Hamiltonian constraint for the two black holes approximate initial data for spinning holes at increasing global finite differences resolution.
  \label{fig:hcnorm2}}
\end{figure}

A study of our perturbative initial data, applied to an equal mass
and equal aligned spins binary
systems with increasing intrinsic spins magnitude, $\chi=S_i/m_i^2$, is
displayed in Fig~\ref{fig:hcnorm2}, showing the control on the violations
of the Hamiltonian constraint during evolution for up to medium magnitudes
of the spins. We also observe the increase
of the levels of violation of the Hamiltonian constraint with the magnitude
of the spins and the lack of reduction of the violations with increasing
global numerical resolution, leading to the conclusion that those violations
are due to the errors produced by the approximation expansion itself.

Specifically, when we try to increase the value of the spins above a perturbative regime  $\chi>0.5$, we see the restrictions of this approach applied to highly spinning holes as displayed in the black and orange lines of Fig.~\ref{fig:hcnorm2} for the case of $\chi=0.8$. This naturally indicates the limitations of the perturbative expansion to represent large values of the spin magnitude.

\begin{figure}
\includegraphics[angle=0,width=.9\columnwidth]{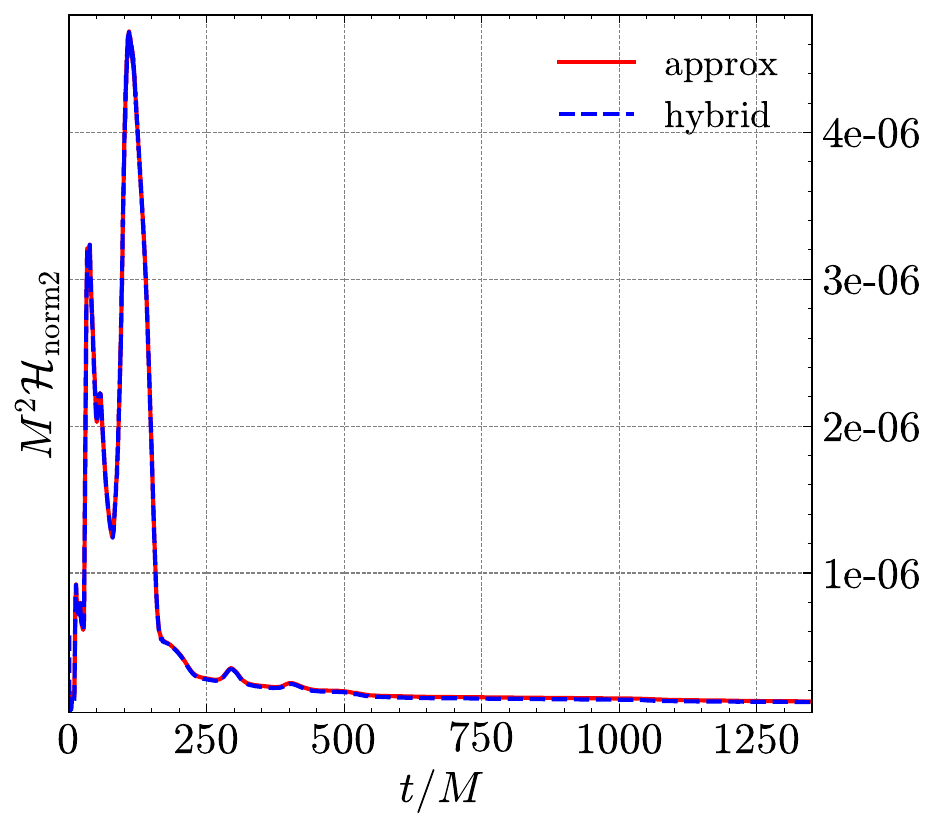}
  \caption{Comparative evolution of the violations of the Hamiltonian constraint for three nonspinning black holes in the full approximation for initial data and for the hybrid exact solution for two black holes and the third approximated.
  \label{fig:3BH1_hcnorm2_comparison}}
\end{figure}
\begin{figure}
\includegraphics[angle=0,width=.9\columnwidth]{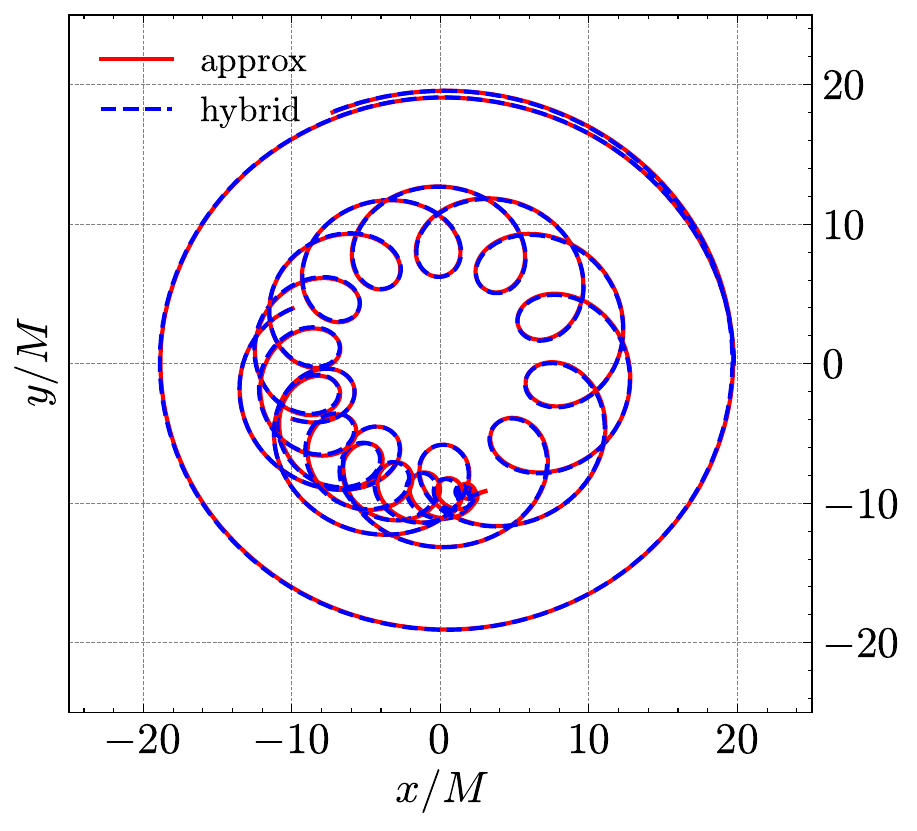}
  \caption{Comparative evolution of the orbital trajectories for three nonspinning black holes in the full approximated initial data and the hybrid exact solution for two black holes plus the third approximated.
  \label{fig:3BH1_orbital_plane_comparison}}
\end{figure}

This suggested us to adopt a hybrid approach to the initial data problem by solving
the highly spinning binary with the usual full numerical approach, ie.
the {\sc TwoPunctures}~\cite{Ansorg:2004ds} code and then superpose
to that solution the third hole in an approximate way. 
We test such approach by reproducing the 3BH1 setup of \cite{Ficarra:2023zjc},
finding consistent results during evolution as displayed in
Fig.~\ref{fig:3BH1_hcnorm2_comparison} for the Hamiltonian constraint
and for the tracks of the three black hole orbits as shown
in Fig.~\ref{fig:3BH1_orbital_plane_comparison}.

We thus conclude that we can use the hybrid approach to the initial
data problem to study evolutions of spinning inner binaries with a third
nonspinning hole in the hierarchical configurations of interest. We note that
full numerical solutions to three black holes initial data
have been studied in Refs.~\cite{Galaviz:2010mx,Bai:2011za,Imbrogno:2021xrh} but
we found our method more practical to implement for exploratory studies.

\section{Three black holes evolutions}\label{sec:3BH}

\begin{figure*}
\begin{center}
\includegraphics[width=1.5\columnwidth]{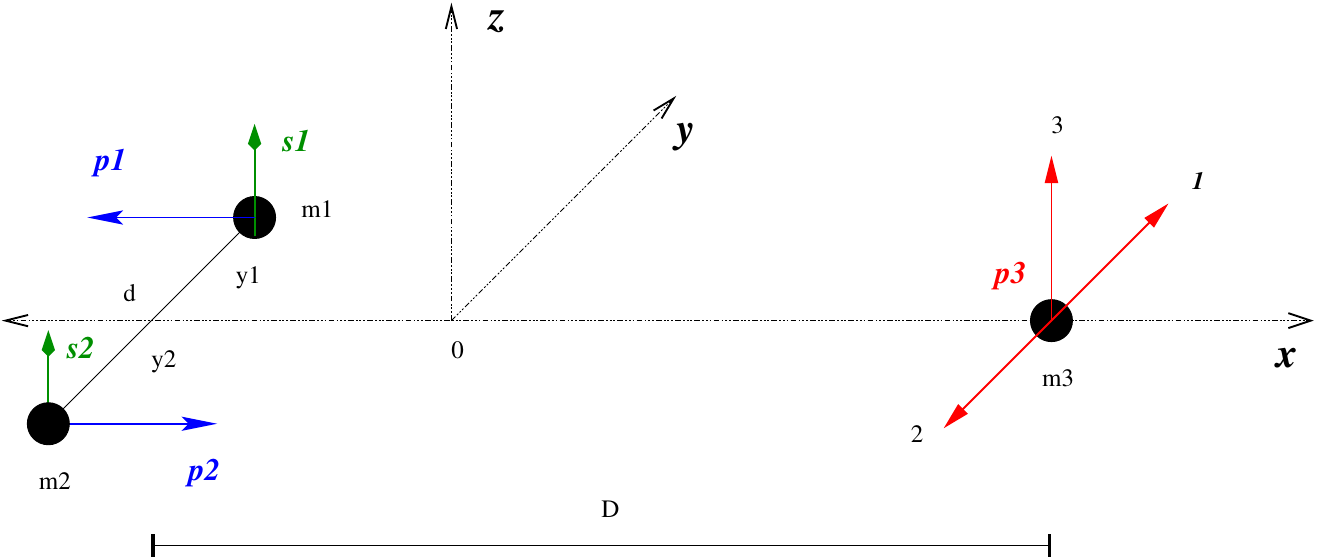}
  \caption{Initial configurations considered
    for the three black hole evolutions, labeled as 3BH1s, 3BH2s, 
for coplanar co- and counter- orbiting, and 3BH3s for polar.}
\label{fig:configs}
\end{center}
\end{figure*}

In this new exploration we will consider a hierarchical prototypical
system with the inner binary at an initial separation of $12m=8M$ and a
third black hole at separation $30M$ (where $m=m_1^H+m_2^H$, the addition
of the horizon masses of the binary, and $M=m_3^H+m$,
the addition of all three horizon masses). All black holes in this first
set will have equal masses (as measured by their individual
horizons) and different relative orbital orientations. Additionally,
the inner binary will have individual spin components $\chi_1 = \chi_2 = +0.8$ along the z-axis. 
This set up is depicted in Fig.~\ref{fig:configs}.
In order to work with small initial eccentricities we consider
the inner binary as isolated and use the
quasicircular formulas of Ref.~\cite{Healy:2017zqj} to obtain the
parameters reported in the first column of Table~\ref{table:spinID-12}
and referred to as 2BH0s. After getting the inner binary parameters we
apply the same quasicircular criteria to 
the outer orbit of the third black hole with a single effective spinning
hole having the added masses and total angular momentum of the
inner binary. This process provides low enough eccentricities
$(e\lesssim0.05)$ for the purposes of our initial study.

\subsection{Hierarchical Three black holes configurations}\label{sec:3BH123}

To start exploring this vast parameter space we have chosen to
consider two coplanar cases, when the third black hole orbit is 
co-orbiting with the binary (3BH1s) and when it is counter-orbiting
(3BH2s). Those parameters are given in Table~\ref{table:spinID-12}.
We also consider a precessing case with the third black hole
momentum perpendicular to the orbital plane of the binary (3BH3s), as
depicted in Fig.~\ref{fig:configs}. In all cases we considered
the quasicircular orbit of the third black hole with the 
inner binary as an effective single black hole as described above.


\input{spin_table_I+II}

\begin{figure}
\includegraphics[angle=0,width=0.9\columnwidth]{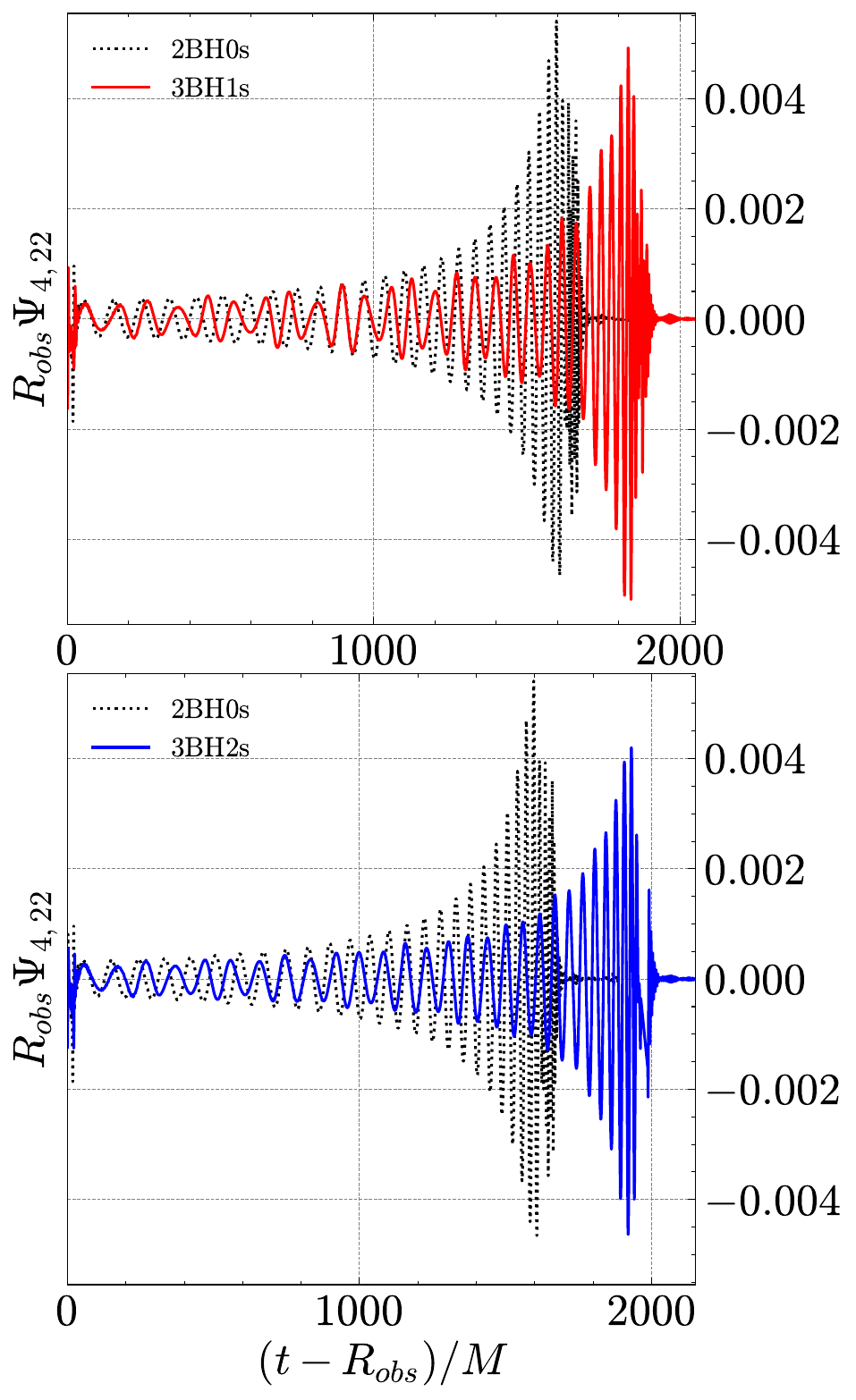}
  \caption{Waveforms generated by the co- and counter- orbiting cases 3BH1s and 3BH2s in
    comparison with the isolated binary 2BH0s. Note the effects of eccentricity
    in the amplitude and post binary merger.
    Here $R_{obs}=113M$ and (2,2) modes extracted.
  \label{fig:isw}}
\end{figure}

\begin{figure}
\includegraphics[angle=0,width=0.9\columnwidth]{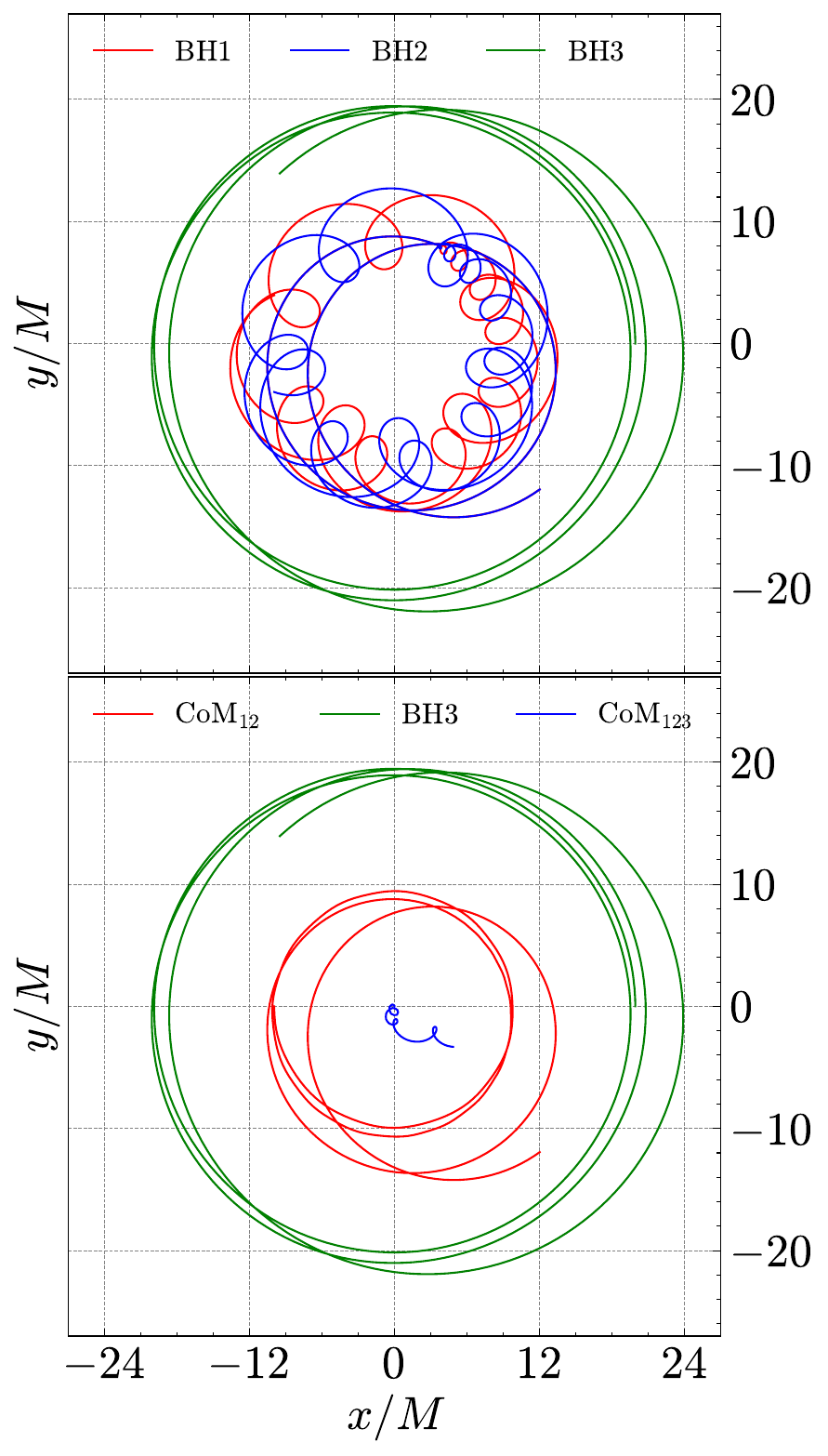}
  \caption{Trajectories of the coplanar co-orbiting case 3BH1s and the evolution of
    the center of masses of the binary and of the three black holes.
    Note the kick after merger.
  \label{fig:is}}
\end{figure}

\begin{figure}
\includegraphics[angle=0,width=0.9\columnwidth]{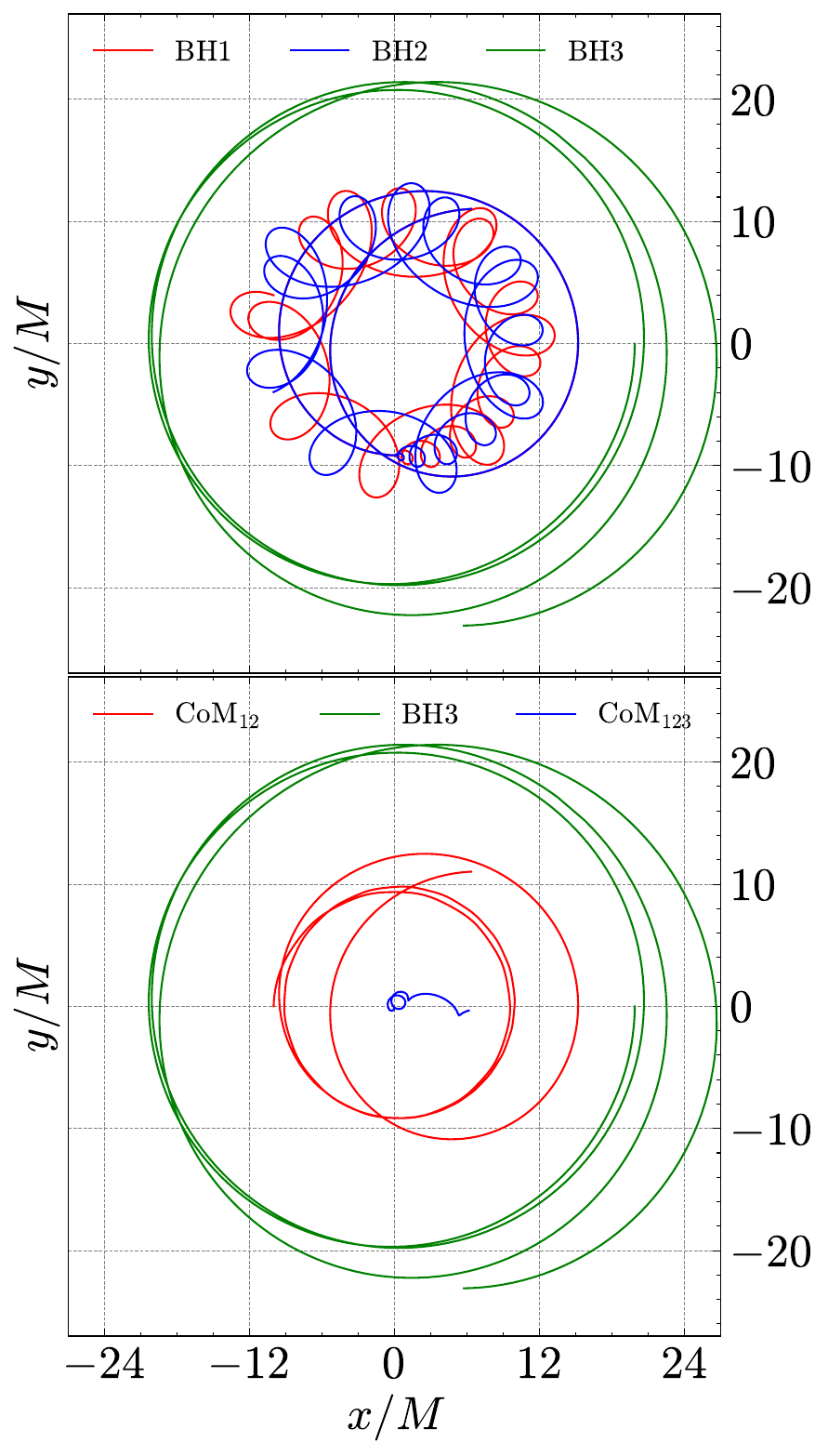}
  \caption{Trajectories of the coplanar counter-orbiting case 3BH2s and the evolution of
    the center of masses of the binary and of the three black holes.
    Note the effects of the kick after merger.
  \label{fig:iis}}
\end{figure}

\begin{figure}
\includegraphics[angle=0,width=0.875\columnwidth]{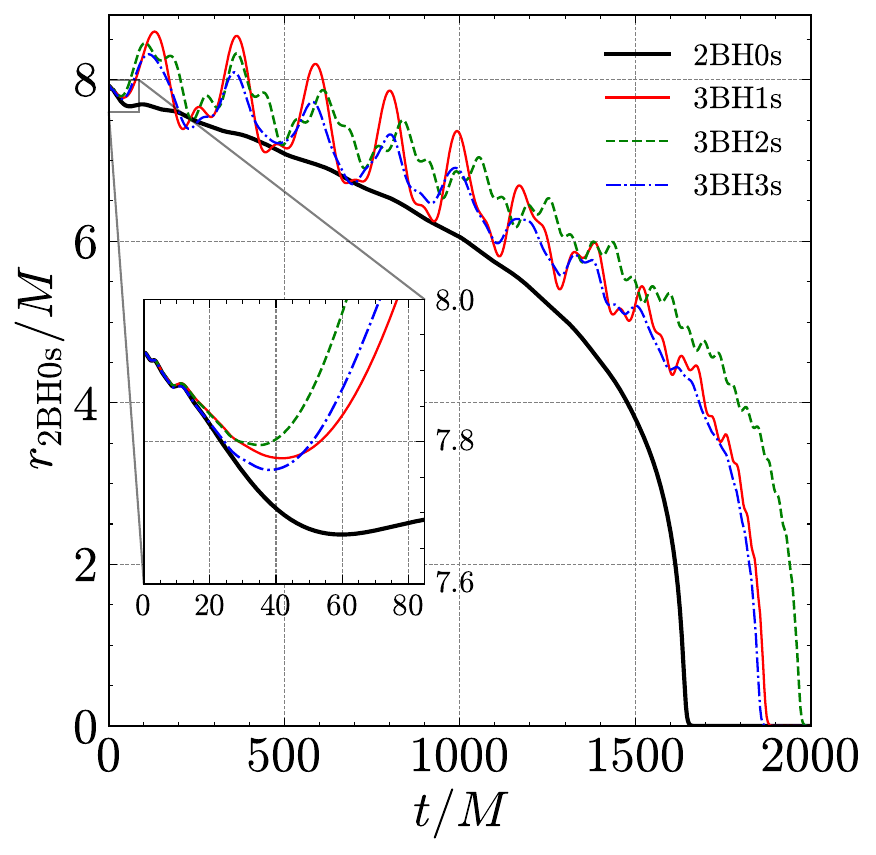}
\caption{Relative separation of the inner binary versus time of the three black hole cases
  studied here 3BH1s-3s in comparison with the isolated binary case. Delayed
		merger and higher eccentricity effects are evident. 
    \label{fig:BH1coordinate}}
\end{figure}

In Fig.~\ref{fig:isw} we display the extracted waveform of
the three black hole simulation 3BH1s. The gravitational
radiation is completely dominated by the inner binary. The
difference with an isolated binary is given by the delay
in the merger due to the presence of the third black hole.
Similar results are obtained for the 3BH2s case, as
seen in the bottom of Fig.~\ref{fig:isw}.
The eccentricity of the 3BH cases
is already apparent in the waveform amplitude versus time
in comparison with the 2BH isolated binary. This eccentricity
effect will also reflect in the tracks of the holes as
we will display below.
Another effect is the orbital motion of the binary and its
merger remnant around the
center of mass of the triple system, as displayed in
Figs.~\ref{fig:is} and \ref{fig:iis}, due to the asymmetric
gravitational radiation of an unequal mass systems.
This displacement leads to a mixing of the gravitational waves modes
as seen at a fixed observer location, but its effects
can be disentangled with techniques like those used
in Refs.~\cite{Woodford:2019tlo,Healy:2020vre}.
The polar case 3BH3s gives a qualitative and quantitative close
picture to these cases, serving as a sort of control of the differences
expected due to precession. Fig.~\ref{fig:BH1coordinate} displays
the comparative
evolution of the inner binary separation in all these three cases.
We note here that a polar and other two precessing cases have been
studied in our previous paper Ref.~\cite{Ficarra:2023zjc} for the nonspinning
binaries and again did not display qualitatively new features, with
results being bracketed between the polar and coplanar cases.

\begin{figure}
\includegraphics[angle=0,width=\columnwidth]{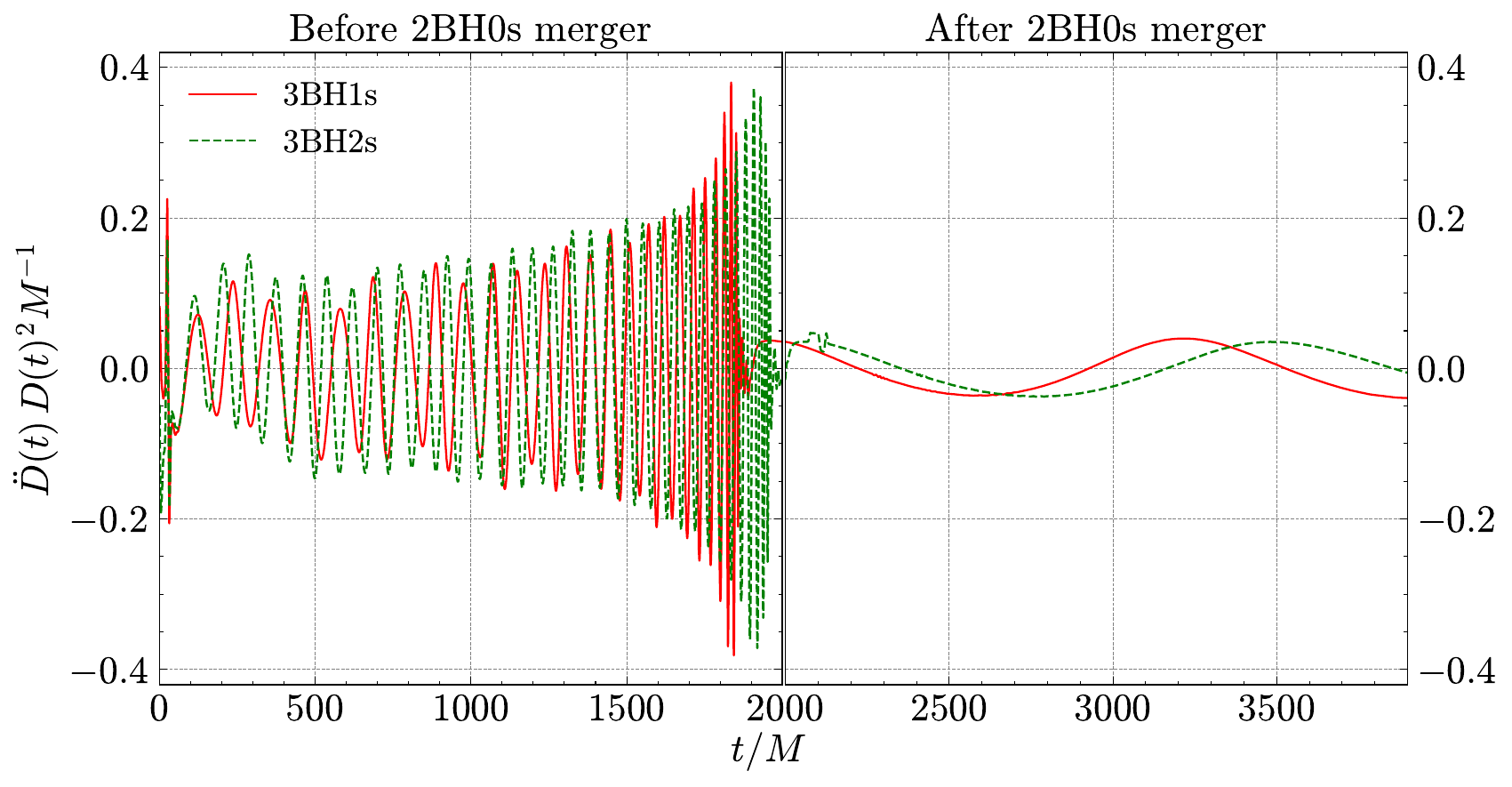}
  \caption{Eccentricity evolution of the outer black hole
  as measured by $D^2\ddot{D}(t)/M$
  for the triple black hole coplanar cases 3BH1s-2s and its
  transition after the inner binary (2BH0s) merger.
    \label{fig:3BHse}}
\end{figure}

The relative motion of the inner binary can be tracked during evolution
of the three black hole system. We can observe comparable initial trajectories
and eccentricities (much larger than in the isolated binary case 2BH0s),
and similar larger merger times for 3BH3s and 3BH1s while the counter-orbiting
3BH2s delays merger by about $\sim100M$.

We can evaluate eccentricity during evolution via the simple formula,
as a function of the separation of the holes, $d$,
$e_d=d^2\ddot{d}/m$, as given in \cite{Campanelli:2008nk}.
We can thus monitor the eccentricity evolution of the orbit of
the third black hole. Before the binary's merger we can refer
the distance of the third hole to the
center of mass of the binary system, as displayed in the 
bottom of Fig.~\ref{fig:is} and ~\ref{fig:iis}, and then after the merger of the
the inner binary to its remnant, as displayed on the right panel of
Fig.~\ref{fig:3BHse}.
We note that the eccentricity measure from the center of mass of the binary
has some total amplitude oscillations with the third black hole orbit and
also seems to grow in time reaching relatively large values before merger.
This seems to be an effect of the use of the coordinates of the
center of mass as a reference for this extended system.
We note that right after merger the eccentricity
measure produces an order of magnitude smaller eccentricity for the
subsequent two orbits of the simulation and produce
values more in line with what we
expect and found for the inner binary studies above.


\subsection{Distance dependence to the third black hole}\label{sec:3BHd}

We next explore how the merger
times and eccentricity evolution of the inner binary vary versus the initial
separation of the outer black hole in our hierarchical setups.
For that end we look again for
quasicircular effective parameters at different initial separations
beyond the reference one at $D=30M$ (relabeled as 3BH1sD0 and 3BH2sD0) 
at increasing initial separations $D/M=35, 40, 50, 60$,
that we label 3BH1sD1-4 and 3BH1sD1-4 
as given in Tables~\ref{table:spinID-3} and~\ref{table:spinID-4}

\input{spin_table_III}
\input{spin_table_IV}

We are interested in studying the effect the third hole has on
the inner binary dynamics. In particular on how it affects the
merger, if prompts or delays it. In Table~\ref{tab:3BH12sD} we
give the results of our simulations versus the initial third
black hole distance to the binary's center of mass. We find
a clear trend towards the delay of the merger, in both quantities,
the merger time and the number of orbits as measured by the
tracks of the holes, where we used as a definition of merger when the
binary distance reaches $d=0.7M$ (which corresponds closely
to the formation of a common horizon).  

\begin{figure}
\includegraphics[angle=0,width=0.9\columnwidth]{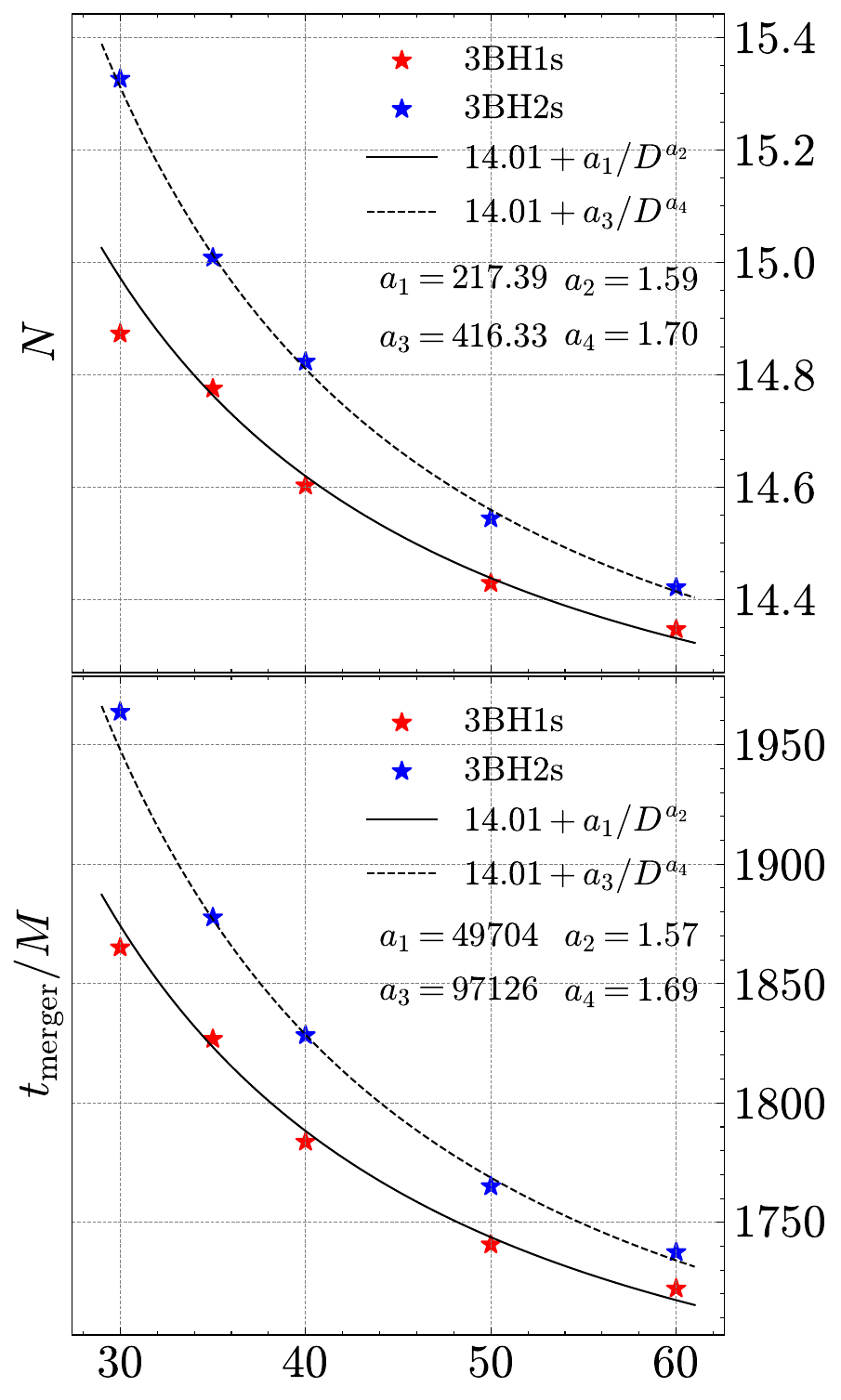}
\caption{Fit to a functional dependence $\rm{2BH0s}+a_{1,3}/D^{a_{2,4}}$
  for the coplanar, co-orbiting 3BHs1D1-4 and counter-orbiting 3BHs2D1-4
  configurations versus distance $D/M$ of the third hole.
  The $D=30M$ points are displayed but not fitted.
  \label{fig:3BHs12d}}
\end{figure}

We model the merger delay as a function of the initial
distance to the third black hole and consider deviations with
respect to the merger time and number of orbits to merger of the
isolated binary, 2BH0s. We then fit a dependence to the data
in Table~\ref{tab:3BH12sD} of the form $\rm{2BH0s}+a_1/D^{a_2}$ for
the four larger separation cases 3BH1,2sD1-4. The
results are displayed in Fig.~\ref{fig:3BHs12d} 
for the third hole co- and counter- orbiting with respect to
the inner binary orbital momentum respectively.
We found in both cases, that a fit leads to
a consistent dependence of the form $\sim1/D^{1.6\pm0.1}$
This is verified when seen in terms of the number of orbits
as well as the evolution time at
the top and bottom of both curves. Note also the consistent
approach of the curves towards a
common value (the isolated binary's) at large separation of the third
hole for both configurations.
Had we included the $D=30$ cases in the fits, we would still find
a -1.6 power dependence, but with larger deviations,
$\sim1/D^{1.6\pm0.2}$.

In Table~\ref{tab:3BH12sD} we report the merger times and number of orbits
for the five cases studied here and the two orbital orientations 1s and 2s.
We first note the clear delay
of the merger of 3BHD0-4 with respect to the isolated binary
2BH0s. We next note the near $1/D^{1.6}$ dependence of the
merger times and number of orbits with initial separation of
the third hole, $D$, and the convergence to the isolated binary
results for large initial separations of the third hole.
We finally note the systematic further merger delay for the counter-orbiting
configuration with respect to the co-orbiting one.

\input{merger_times_spin_table}


\section{Other three black hole configurations}\label{sec:3BHSandQ}

In this section we explore two other 3BH configurations of potential
astrophysical interest,
the case of a third black hole in a close scattering trajectory with
respect to a binary, and the case of a binary orbiting a 
more massive third black hole.

\subsection{Scattering Three black holes configurations}\label{sec:3BHS}

We have seen that our studies of three black hole evolutions
indicate some evidence that those three body interactions seem to
be dominated by the closest or 'sudden' approach of one component of the
inner binary to the third hole. In order to further explore this
hypothesis we can try to single out this 'sudden' effect by studying
configurations where the third black hole approaches the binary in
a close scattering orbit. 

\begin{figure}
\includegraphics[angle=0,width=.9\columnwidth]{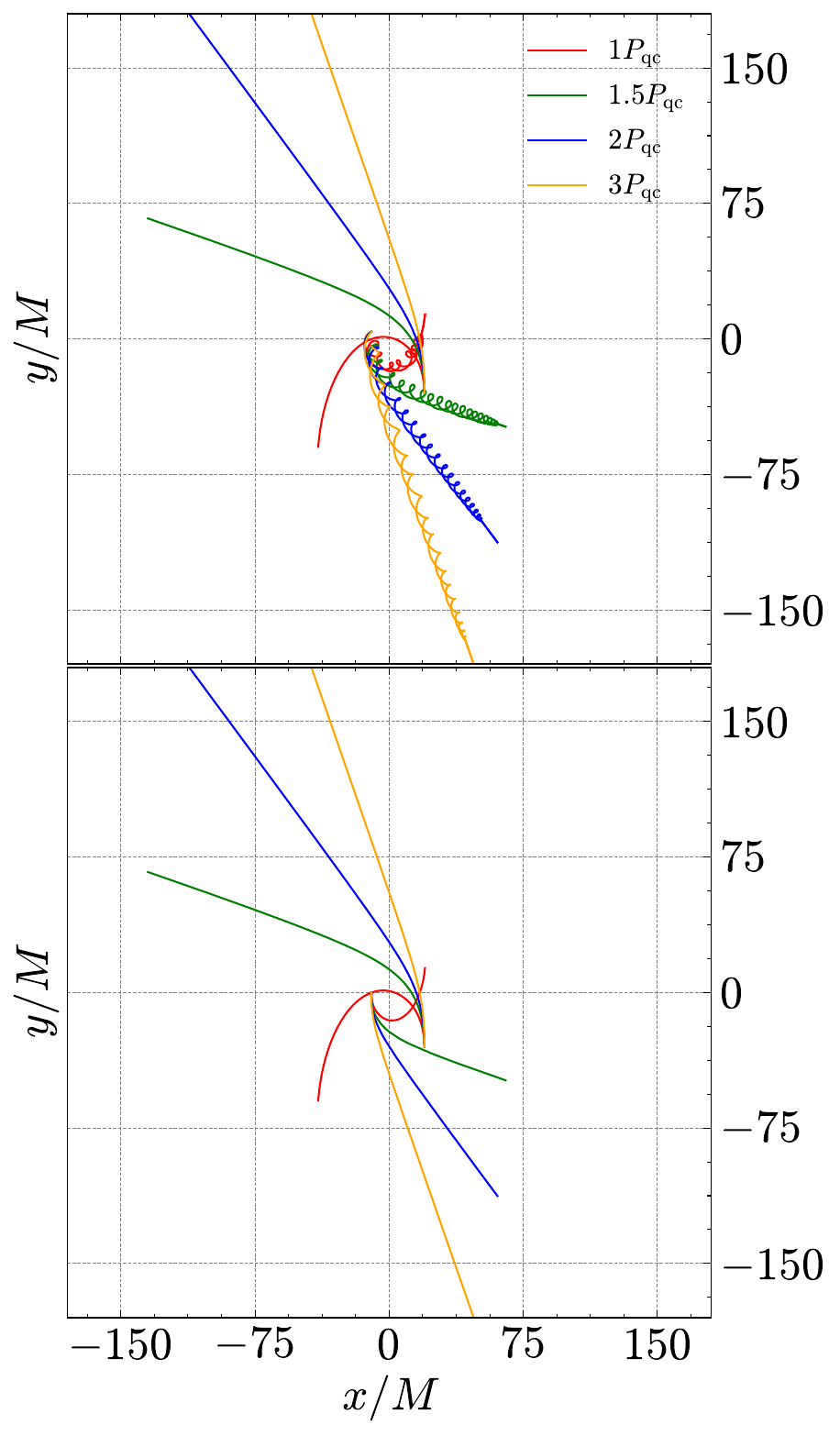}
  \caption{Trajectories of the binary until merger with the third black hole starting at the same distance $D=42M$ but with increasing linear momentum used for the quasicircular previous cases by factors $\times 1, 1.5, 2, 3$. Below the trajectory of the binary's center of mass and the third black hole. 
  \label{fig:3BHscattering}}
\end{figure}

In order to simplify the analysis and to
relate the scattering orbits to the previous quasicircular
hierarchical systems in \cite{Ficarra:2023zjc}
we will come back to the equal mass nonspinning
systems and further displace the initial location of the
third hole to coordinates $(x,y)=(30M,30M)$ with respect to the
center of mass of the inner binary. We have also ensured to have a set
of simulations with increasing
magnitude of the linear momentum from the quasicircular $P_{\rm{qc}}$ at that
$d=42M$ location by using a factor $\times(1, 1.5, 2, 3)$ of the
original $P_{qc}(d=30M)$,
as described in Table~\ref{table:scatteringID}, and labeled them as
3BHscat1-4 respectively.
Using as a reference the factor $1-f=P_T/P_{\rm{qc}}$ of
the actual tangential linear momentum to the corresponding quasicircular
one \cite{Ciarfella:2022hfy}, in the Newtonian approximation $f<1-\sqrt{2}$
would lead to scattering orbits. Thus our first choice would lead to
an elliptic orbit and the other three to scattering orbits as we observe
in the actual full numerical tracks displayed in Figure~\ref{fig:3BHscattering},
where we also observe the triggering of the binary's merger at successive
later times. The precise merger times are reported in Table~\ref{tab:3BHS}
and they indicate a correlation between the merger times (corrected by the
time of closest approach $t_{min}$) with the closest distance reached by
the scattering hole to the binary. The closest the approach, the later
the merger 
(except for the first case $1P_{\rm{qc}}$ that leads to a bound elliptic
orbit and that at such closest approach prompts an early merger as seen
in our former studies \cite{Lousto:2007rj} of close encounters of three
black holes).


\input{scattering_table}

\input{merger_times_scattering_table}

This dependence with the closest approach, even for fast moving black holes,
somewhat confirm the relevance of the concept of a 'sudden' interaction of
third black holes with an inner binary and may be used to develop approximation
techniques, for instance of the sort of \cite{Rettegno:2023ghr} and references
therein.

\subsection{High mass ratio Three black holes configurations}\label{sec:3BHq}

\begin{figure}
\includegraphics[angle=0,width=0.9\columnwidth]{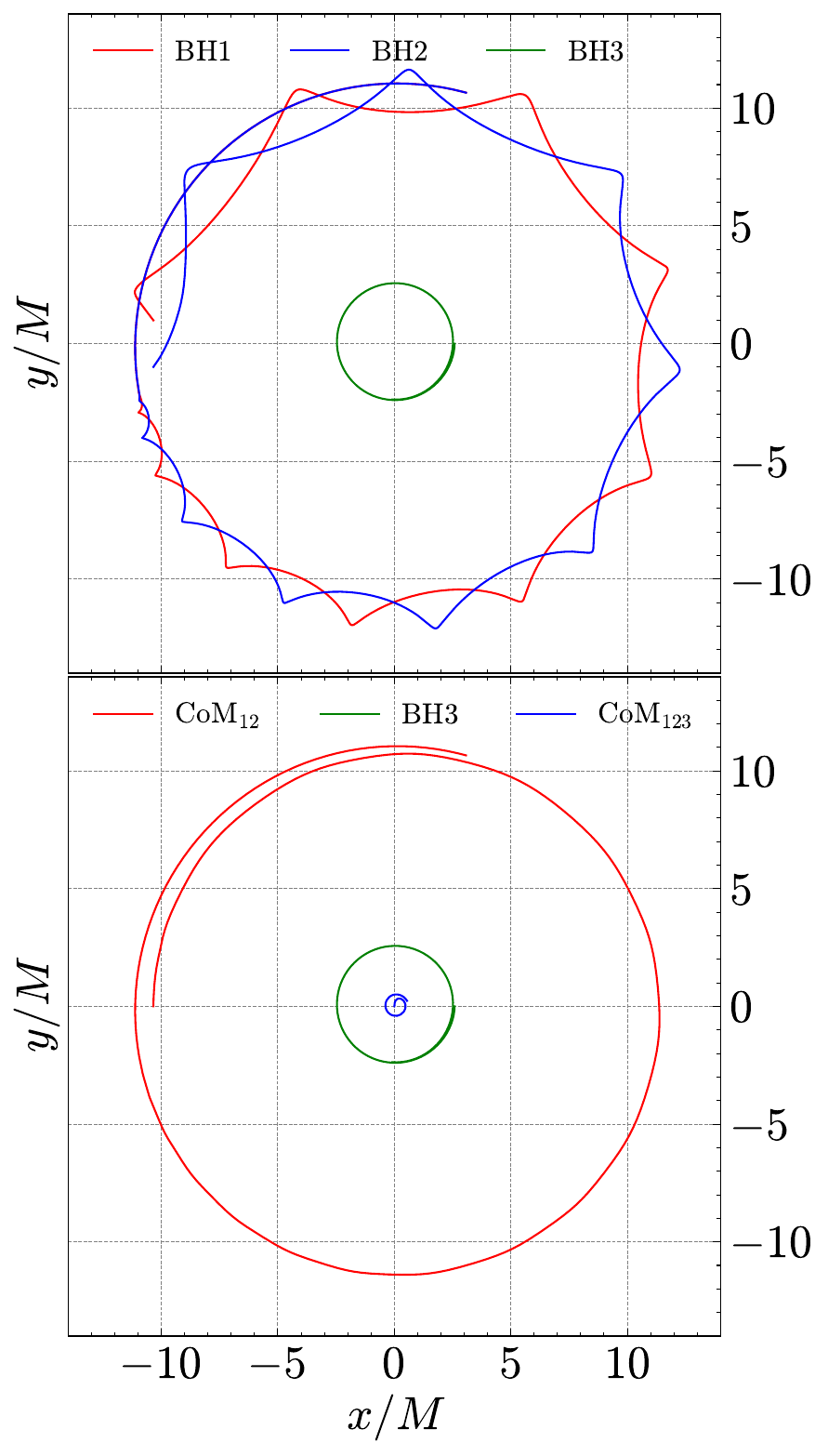}
  \caption{The hierarchical orbits of the three black holes in the center of mass frame (see also lower panel) for the 8:1:1 unequal mass 3BH2q8 configuration. The binary center of mass is moving clockwise in the figure while the binary itself the opposite.
  \label{fig:3BH2q8_CofM}}
\end{figure}

Interesting astrophysical scenarios involving three black holes involve
stellar mass binary interactions in globular clusters \cite{Gultekin:2003xd}
with intermediate mass holes at its core \cite{Miller:2003sc}.
These interactions may lead to eccentricity increase in the binary
\cite{DallAmico:2023neb} with implications for the characteristics of the 
gravitational waves \cite{Pina:2023vjv}.
In order to start exploring these effects during the latest stages of
the binary, previous to its merger, we will consider a sequence of increasingly
dispair mass ratios of the binary (which itself is chosen formed by
equal mass nonspinning holes) to a much more massive third hole,
modeling an intermediate mass black hole
at the core of a globular cluster. We thus start comparing mass ratios a factor eight to each
the binary's individual mass black holes 8:1:1 and then raise it to an eighteen mass ratio
18:1:1 as displayed in Table \ref{table:massratioID}.
We considered coplanar orbits prograde and retrograde with respect to the
larger hole labeled as 3BH1q and 3BH2q respectively.
Figure~\ref{fig:3BH2q8_CofM} displays the orbital motion of the three holes as seen
in the center of mass frame for the 3BH2q8 configuration. The binary merges at
about completing an orbit around the larger hole. The bottom plot tracks
the center of mass of the binary, with an expected follow up slow decay around the larger hole (and eccentricity),
product of the gravitational radiation mostly at low frequencies, while the merging of
the binary itself contributing mostly at higher gravitational waves frequencies
(although, of course, the three body system radiates as a whole).

\begin{figure}
\includegraphics[angle=0,width=.9\columnwidth]{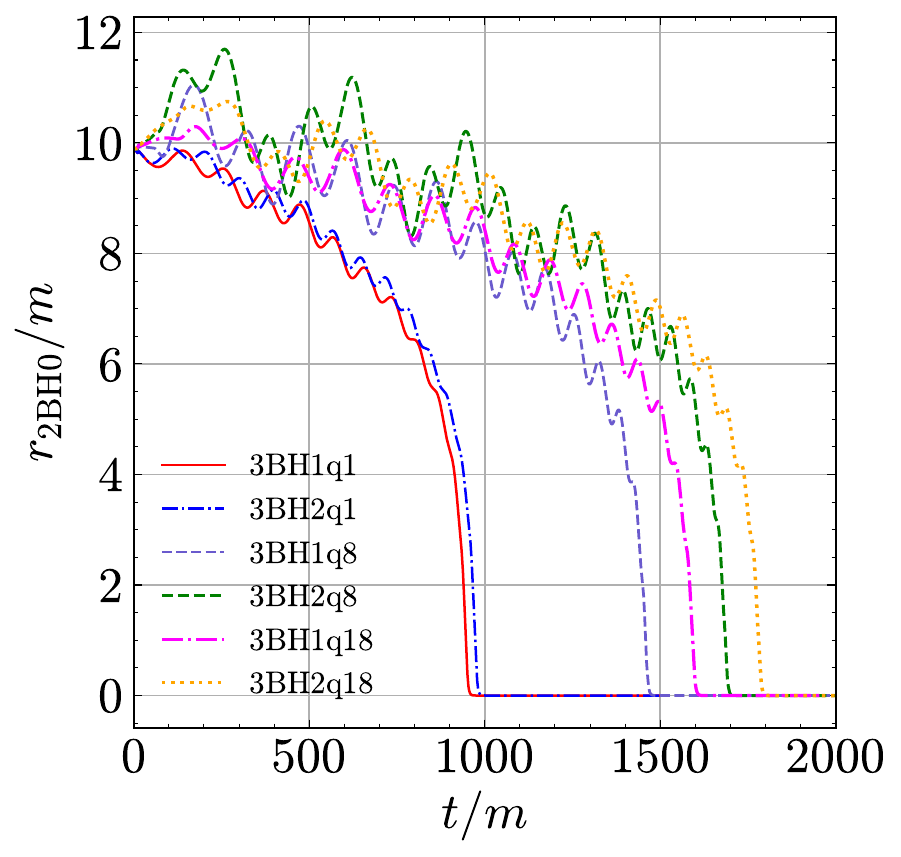}
\caption{The time evolution of the binary's separation $d=r_{2BH0}/m$ displays the eccentricity effects triggered by the third (more massive) black hole and the merger delay due to the smaller mass ratio and the orientation of the orbit. 
  \label{fig:3BH2q8_separation}}
\end{figure}

The sequence of increasing mass ratios can be followed, 
but we found the trends of binary's merger times to be already clear.
In fact, Figure \ref{fig:3BH2q8_separation} displays a pattern of increased
merger time with the increase of the mass ratio, as expected due to the
lower efficiency of gravitational radiation relative to the total mass of the system 
with respect to the equal mass cases 3BH1q1 and 3BH2q1
(that we also include in the study to serve as a reference).
We also note the relative delay for each of the mass ratios between
their corresponding prograde and retrograde configurations. Notably
the retrograde cases delay their merger with respect to the prograde
possibly due to its shortest duration close approach interactions with the
larger third black hole. An effect further supporting the notion
of the 'sudden' interaction predominance in three black hole configurations.
Precise times and number of orbits to merger are compiled in
Table~\ref{tab:3BHsq}. We also note here that this reverses the sense
of the hangup effect \cite{Campanelli:2006uy,Healy:2018swt}
observed in binary systems with spins, where spins aligned with the
orbital angular momentum lead to a merger delay and antialigned spins
to a prompt merger.

\input{massratio_table}

\input{merger_times_massratio_table}

\section{Conclusions and Discussion}\label{sec:Discussion}

Our full analytic approach to solve perturbatively (Bowen-York)
initial data for
three black holes is naturally limited by the smallness of the
expansion parameters, namely spins and linear momenta of the holes,
as well as the inverse of its initial separation. While the last two
parameters are often times small, the extension to highly spinning black
holes shows inaccuracies.
To cure those we have implemented a hybrid approach solving the inner
binary with high spins fully numerically, in the usual way with the
TwoPunctures solver of the Hamiltonian constraint, and then add up
the analytic expansion of the third hole contribution.
We validated this hybrid procedure in the small spin regime and
the whole procedure proved accurate enough for our
current exploratory studies.
We also refer to the convergence studies and more details of the
analytic and 
numerical techniques to the previous paper \cite{Ficarra:2023zjc}
in this series.

We then revisited the triple black hole scenario to study their
merging times and eccentricity evolution in the presence of spins.
We found that the third
black hole delays the merger of the spinning binary by
an inverse a power of the distance, $\sim1/D^{1.6}$. This
behavior is clearly different to those of the nonspinning configurations
studied in our previous work \cite{Ficarra:2023zjc}, that 
was $\sim1/D^{2.5}$. While we interpreted the nonspinning three black
hole interactions dominated by tidal effects, now in the spinning case
we interpret this leading interaction as being dominated by a sort
of spin-orbit effect (scaling like $\sim1/D^{1.5}$), taking place
during the closest approach and for a short term, namely
a ``sudden'' interaction
of the third hole with the closest hole of the binary at that moment.
We also observe here that if we start the third black hole
closer to $\approx30M$ this leads to a prompt disruption and a
binary or merger as observed in previous work \cite{Lousto:2007rj,Ficarra:2023zjc}.

This interesting new dependence of the merger time with the distance
of the third hole to the binary in the presence of the spin interactions
prompted the study of scattering configurations of the third hole to
study its effects on triggering merger and eccentricity on inner
binaries to check the idea of the ``sudden'' interaction as the
leading effect.
In fact, we verified a dependence of the merger times with the
scattering closest approach distance to the binary as described in
Sec.~\ref{sec:3BHS}, even for nonspinning holes. These effects have
the potential to display distinctive features, for instance, 
gravitational waves from the scattering of two black holes have
been studied in \cite{OLeary:2008myb} and its detectability in
\cite{Mukherjee:2020hnm}.

Finally, we have also studied the mass ratio dependence on the
inner binary orbiting a much larger third black hole. This could
be a prototypical study for a stellar masses black hole binary
near a central intermediate mass hole in a globular cluster.
Our explicit studies are for mass
ratios 8:1:1 and 18:1:1, but the sequence could be continued for
at least until 128:1:1 systems, since we have proven our full numerical
evolution techniques work very well \cite{Lousto:2020tnb}  applied to binary
black holes holding such small mass ratios.


\begin{acknowledgments}
The authors thank Alessandro Ciarfella, James Healy, Hiroyuki Nakano, and
Yosef Zlochower for useful discussions. The authors also gratefully acknowledge
the National Science Foundation (NSF) for financial support from Grant
PHY-2207920.  Computational resources were also
provided by the Blue Sky, Green Prairies, and White
Lagoon clusters at the CCRG-Rochester Institute of Technology, which
were supported by NSF grants 
No.\ AST-1028087, No.\ PHY-1229173, No.\ PHY-1726215, and
No.\ PHY-2018420.  This work used the ACCESS allocation TG-PHY060027N,
founded by NSF, and project PHY20007 Frontera, an NSF-funded Petascale
computing system at the Texas Advanced Computing Center (TACC).
\end{acknowledgments}

\bibliographystyle{apsrev4-1}
\bibliography{../../../Bibtex/references.bib}

\end{document}

%% file: spin_table_I+II.tex
\begin{table}[ht]
\centering
\caption{Initial data parameters for the base binary (2BH0s), two coplanar (3BH1s, 3BH2s) and polar (3BH3s) configurations with a third black hole at a distance $D$ from the binary along the $x$-axis. $(x_i,y_i,z_i)$ and $(p^x_i,p^y_i,p_i^z)$ are the initial position and momentum of the puncture $i$, $m^p_i$ is the puncture mass parameter, $m^H_i$ and $\chi^H_{iz}$ are the horizon mass and dimensionless spin at $t = 0$, $M\Omega$ is the binary’s orbital frequency, $d$ is the binary's initial coordinate separation and $d_{spd}$ is the binary's proper distance. Parameters not specified are zero.}
\label{table:spinID-12}
\begin{tabular}{lcccc}
\toprule
         Config &       2BH0s &       3BH1s &       3BH2s &       3BH3s \\
\hline
        $x_1/M$ & -9.94446878 & -9.94446878 & -9.99075115 & -9.96704932 \\
        $y_1/M$ &  3.96250893 &  3.96250893 &  3.96250893 &  3.96250893 \\
        $z_1/M$ &  0.00000000 &  0.00000000 &  0.00000000 &  0.00000000 \\
      $p_1^x/M$ & -0.05463934 & -0.05462799 & -0.05462652 & -0.05462799 \\
      $p_1^y/M$ & -0.03157750 & -0.02172902 &  0.02175594 & -0.00031577 \\
      $p_1^z/M$ &  0.00000000 &  0.00000000 &  0.00000000 & -0.02169541 \\
      $m_1^p/M$ &  0.20246647 &  0.20012400 &  0.20010400 &  0.20010600 \\
      $m_1^H/M$ &  0.33147290 &  0.33152795 &  0.33152152 &  0.33152582 \\
$\chi_{1z}^H/M$ &  0.80906581 &  0.80880423 &  0.80883273 &  0.80880178 \\
        $x_2/M$ & -9.94446878 & -9.94446878 & -9.99075115 & -9.96704932 \\
        $y_2/M$ & -3.96250893 & -3.96250893 & -3.96250893 & -3.96250893 \\
        $z_2/M$ &  0.00000000 &  0.00000000 &  0.00000000 &  0.00000000 \\
      $p_2^x/M$ &  0.05463934 &  0.05465069 &  0.05465216 &  0.05708188 \\
      $p_2^y/M$ &  0.03157750 & -0.02109747 &  0.02238748 &  0.00036813 \\
      $p_2^z/M$ &  0.00000000 &  0.00000000 &  0.00000000 & -0.02166824 \\
      $m_2^p/M$ &  0.20246647 &  0.20012400 &  0.20010400 &  0.20010600 \\
      $m_2^H/M$ &  0.33147321 &  0.33152655 &  0.33152212 &  0.33152648 \\
$\chi_{2z}^H/M$ &  0.80906785 &  0.80881193 &  0.80881481 &  0.80880501 \\
          $d/M$ &  7.92501785 &  7.92501785 &  7.92501785 &  7.92501785 \\
    $d_{spd}/M$ &  11.0475159 &  11.1643519 &  11.1545122 &  11.1645980 \\
        $x_3/M$ &             &  19.9492516 &  19.9054250 &  19.9277120 \\
        $y_3/M$ &             &  0.00000000 &  0.00000000 &  0.00000000 \\
        $z_3/M$ &             &  0.00000000 &  0.00000000 &  0.00000000 \\
      $p_3^x/M$ &             & -0.00002269 & -0.00002564 & -0.00002269 \\
      $p_3^y/M$ &             &  0.04282649 & -0.04414343 &  0.00000000 \\
      $p_3^z/M$ &             &  0.00000000 &  0.00000000 &  0.04339083 \\
      $m_3^p/M$ &             &  0.32906500 &  0.32902500 &  0.32902500 \\
      $m_3^H/M$ &             &  0.33332869 &  0.33333254 &  0.33330730 \\
      $M\Omega$ & 0.032314209 &  0.00581907 &  0.00587723 &  0.00583637 \\
          $D/M$ &             &  29.8937203 &  29.8961761 &  29.8947610 \\
\hline
\end{tabular}
\end{table}

%% file: spin_table_III.tex
\begin{table}
\centering
\caption{Initial data parameters for coplanar configurations with a third black hole placed at different distances $D$ from the binary along the $x$-axis, 3BH1sD1-4.}
\label{table:spinID-3}
\begin{tabular}{lccccc}
\toprule
         Config &     3BH1sD1 &     3BH1sD2 &     3BH1sD3 &     3BH1sD4 \\
\hline
        $x_1/M$ &  -11.612696 &  -13.280575 & -16.6156958 & -19.9503036 \\
        $y_1/M$ &  3.96250893 &  3.96250893 &  3.96250893 &  3.96250893 \\
      $p_1^x/M$ & -0.05463218 & -0.05463453 & -0.05463687 & -0.05463791 \\
      $p_1^y/M$ & -0.02000458 & -0.01863566 & -0.01657699 & -0.01508252 \\
      $m_1^p/M$ &  0.20052900 &  0.20080000 &  0.20128000 &  0.20145000 \\
      $m_1^H/M$ &  0.33154561 &  0.33154757 &  0.33158544 &  0.33155661 \\
$\chi_{1z}^H/M$ &  0.80869972 &  0.80868571 &  0.80850782 &  0.80863483 \\
        $x_2/M$ &  -11.612696 &  -13.280575 & -16.6156958 & -19.9503036 \\
        $y_2/M$ & -3.96250893 & -3.96250893 & -3.96250893 & -3.96250893 \\
      $p_2^x/M$ &  0.05464651 &  0.05464415 &  0.05464181 &  0.05464078 \\
      $p_2^y/M$ & -0.01937303 & -0.01800411 & -0.01594544 & -0.01445097 \\
      $m_2^p/M$ &  0.20052900 &  0.20080000 &  0.20128000 &  0.20145000 \\
      $m_2^H/M$ &  0.33154384 &  0.33154772 &  0.33158381 &  0.33155726 \\
$\chi_{2z}^H/M$ &  0.80870507 &  0.80870384 &  0.80851186 &  0.80865377 \\
          $d/M$ &  7.92501785 &  7.92501785 &  7.92501785 &  7.92501785 \\
    $d_{spd}/M$ &  11.1475754 &  11.1350617 &  11.1172208 &  11.1056409 \\
        $x_3/M$ &  23.2803971 &  26.6120360 & 33.27622481 & 39.94114485 \\
        $y_3/M$ &  0.00000000 &  0.00000000 &  0.00000000 &  0.00000000 \\
      $p_3^x/M$ & -0.00001433 & -0.00000962 & -0.00000495 & -0.00000287 \\
      $p_3^y/M$ &  0.03937762 &  0.03663976 &  0.03252242 &  0.02953350 \\
      $m_3^p/M$ &  0.32965000 &  0.33009500 &  0.33120500 &  0.33120500 \\
      $m_3^H/M$ &  0.33331060 &  0.33330102 &  0.33331759 &  0.33334524 \\
      $M\Omega$ &  0.00464832 &  0.00382336 & 0.002754558 & 0.002104978 \\
          $D/M$ &  34.8930391 &  39.8926110 &  49.8919206 & 59.89144850 \\
\hline
\end{tabular}
\end{table}

%% file: spin_table_IV.tex
\begin{table}
\centering
\caption{Initial data parameters for coplanar configurations with a third black hole placed at different distances $D$ from the binary along the $x$-axis, 3BH2sD1-4.}
\label{table:spinID-4}
\begin{tabular}{lccccc}
\toprule
         Config &     3BH2sD1 &     3BH2sD2 &     3BH2sD3 &      3BH2sD4 \\
\hline
        $x_1/M$ & -11.6543899 & -13.3187754 & -16.6488714 &  -19.9799919 \\
        $y_1/M$ &  3.96250893 &  3.96250893 &  3.96250893 &   3.96250893 \\
      $p_1^x/M$ & -0.05463145 & -0.05463414 & -0.05463673 &  -0.05463784 \\
      $p_1^y/M$ &  0.01985426 &  0.01837108 &  0.01617898 &   0.01461254 \\
      $m_1^p/M$ &  0.20053000 &  0.20080000 &  0.20128000 &   0.20145000 \\
      $m_1^H/M$ &  0.33154688 &  0.33154740 &  0.33158433 &   0.33155782 \\
$\chi_{1z}^H/M$ &  0.80870397 &  0.80868774 &  0.80850861 &   0.80865279 \\
        $x_2/M$ & -11.6543899 & -13.3187754 & -16.6488714 &  -19.9799919 \\
        $y_2/M$ & -3.96250893 & -3.96250893 & -3.96250893 &  -3.96250893 \\
      $p_2^x/M$ &  0.05464723 &  0.05464454 &  0.05464195 &   0.05464084 \\
      $p_2^y/M$ &  0.02048581 &  0.01900263 &  0.01681053 &   0.01524409 \\
      $m_2^p/M$ &  0.20053000 &  0.20080000 &  0.20128000 &   0.20145000 \\
      $m_2^H/M$ &  0.33154769 &  0.33154828 &  0.33158463 &   0.33156030 \\
$\chi_{2z}^H/M$ &  0.80868914 &  0.80867610 &  0.80849371 &   0.80865465 \\
          $d/M$ &  7.92501785 &  7.92501785 &  7.92501785 &   7.92501785 \\
    $d_{spd}/M$ & 11.14783586 & 11.12480118 & 11.10687866 & 11.105719020 \\
        $x_3/M$ &  23.2406336 &  26.5754054 &  33.2441618 &   39.9122978 \\
        $y_3/M$ &  0.00000000 &  0.00000000 &  0.00000000 &   0.00000000 \\
      $p_3^x/M$ & -0.00001577 & -0.00001040 & -0.00000523 &  -0.00000299 \\
      $p_3^y/M$ & -0.04034007 & -0.03737370 & -0.03298950 &  -0.02985663 \\
      $m_3^p/M$ &  0.32965000 &  0.33009500 &  0.33075000 &   0.33120500 \\
      $m_3^H/M$ &  0.33334044 &  0.33332222 &  0.33332950 &   0.33335266 \\
      $M\Omega$ &  0.00468542 & 0.003848442 & 0.002767571 &  0.002112571 \\
          $D/M$ &  34.8950235 &  39.8941808 &  49.8930332 &   59.8922897 \\
\hline
\end{tabular}
\end{table}

%% file: merger_times_spin_table.tex
\begin{table}
\centering
\caption{Number of orbits to merger and merger time of the inner binary for different orbital distances of the third black hole. Cases 3BHD0-4 where (1s) is prograde and (2s) retrograde orbits.}
\label{tab:3BH12sD}
\begin{ruledtabular}
\begin{tabular}{lllll}
$D/M$ & $\#$orbits & $t_{\text{merger}}/M$ & $\#$orbits & $t_{\text{merger}}/M$\\
  & (1s) & (1s) & (2s) & (2s)\\
  \hline
   30 &      14.87 &              1865.10 &      15.33 &              1963.75\\
   35 &      14.78 &              1826.77 &      15.00 &              1877.71\\
   40 &      14.60 &              1783.75 &      14.82 &              1828.33\\
   50 &      14.43 &              1740.73 &      14.54 &              1765.10\\
   60 &      14.35 &              1722.29 &      14.42 &              1737.50\\
   $\infty$ &      14.01 &              1638.02 &      14.01 &              1638.02 \\
\end{tabular}
\end{ruledtabular}
\end{table}

%% file: scattering_table.tex
\begin{table}
\centering
\caption{Initial parameters for equal-mass, non-spinning scattering configurations.}
\label{table:scatteringID}
\begin{tabular}{lcccc}
\toprule
     Config &     3BHscat1 &    3BHscat2 &    3BHscat3 &    3BHscat4 \\
\hline
    $x_1/M$ &  -9.95027835 & -9.95027835 & -9.95027835 & -9.95027835 \\
    $y_1/M$ &   3.96250893 &  3.96250893 &  3.96250893 &  3.96250893 \\
  $p_1^x/M$ &  -0.05705839 & -0.05705839 & -0.05705839 & -0.05705839 \\
  $p_1^y/M$ &  -0.02179566 & -0.03250943 & -0.04322319 & -0.06465073 \\
  $m_1^p/M$ &   0.32406500 &  0.32392700 &  0.32352700 &  0.32272500 \\
  $m_1^H/M$ &   0.33323974 &  0.33332045 &  0.33321649 &  0.33326252 \\
    $x_2/M$ &  -9.95027835 & -9.95027835 & -9.95027835 & -9.95027835 \\
    $y_2/M$ &  -3.96250893 & -3.96250893 & -3.96250893 & -3.96250893 \\
  $p_2^x/M$ & 0.0570813769 &  0.05708138 &  0.05708138 &  0.05708138 \\
  $p_2^y/M$ &  -0.02105940 & -0.03177317 & -0.04248694 & -0.06391447 \\
  $m_2^p/M$ &   0.32406500 &  0.32392700 &  0.32352700 &  0.32272500 \\
  $m_2^H/M$ &   0.33339854 &  0.33347563 &  0.33336817 &  0.33340831 \\
      $d/M$ &   7.92501786 &  7.92501786 &  7.92501786 &  7.92501786 \\
$d_{spd}/M$ & 10.623675275 & 10.63025694 & 10.63841909 & 10.66317740 \\
    $x_3/M$ &   19.7641253 &  19.7641253 &  19.7641253 &  19.7641253 \\
    $y_3/M$ &  -30.0000000 & -30.0000000 & -30.0000000 & -30.0000000 \\
  $p_3^x/M$ &  -0.00002299 & -0.00002299 & -0.00002299 & -0.00002299 \\
  $p_3^y/M$ &   0.04285506 &  0.06428260 &  0.08571013 &  0.12856519 \\
  $m_3^p/M$ &   0.33008000 &  0.32925000 &  0.32815000 &  0.32506000 \\
  $m_3^H/M$ &   0.33331900 &  0.33330898 &  0.33333513 &  0.33332132 \\
\hline
\end{tabular}
\end{table}

%% file: merger_times_scattering_table.tex
\begin{table}
\centering
\caption{Merger time, time and distance of closest approach of the inner binary for different values of the initial linear momentum of the third black hole.}
\label{tab:3BHS}
\begin{ruledtabular}
\begin{tabular}{llll}
Configuration & $t_{\text{merger}}/M$ & $t_{\text{min}}/M$ & $D_{\text{min}}/M$ \\
\hline
   1$P_{\rm{qc}}$ &      686.94  &      240.19 &    15.60 \\
   1.5$P_{\rm{qc}}$ &    1222.81 &      164.31 &    23.82\\
   2$P_{\rm{qc}}$ &      1150.69 &      116.00 &    26.41\\
   3$P_{\rm{qc}}$ &      1072.50 &      77.90  &    27.93\\
\end{tabular}
\end{ruledtabular}
\end{table}

%% file: massratio_table.tex
\begin{table*}
\centering
\caption{Initial parameters for non-spinning coplanar setups with mass ratios $m_3^H:m_2^H:m_1^H$ of 1:1:1, 8:1:1 and 18:1:1.}
\label{table:massratioID}
\begin{tabular}{lcccccc}
\toprule
     Config &      3BH1q1 &       3BH2q1 &       3BH1q8 &       3BH2q8 &     3BH1q18 &     3BH2q18 \\
\hline
    $x_1/M$ & -9.95121484 &  -9.98358508 &  -10.3327592 &   -10.341799 &  -10.758263 &  -10.760757 \\
    $y_1/M$ &  3.29853493 &   3.29853493 &   0.98956048 &   0.98956048 &  0.49478024 &  0.49478024 \\
  $p_1^x/M$ & -0.06462839 &  -0.06462736 &  -0.01930654 &  -0.01930063 & -0.00965915 & -0.00965788 \\
  $p_1^y/M$ & -0.02218703 &   0.02128140 &  -0.02595764 &   0.02601309 & -0.01544532 &  0.01538831 \\
  $m_1^p/M$ &  0.32205500 &   0.32205500 &   0.09380000 &   0.09380000 &  0.04650000 &  0.04650000 \\
  $m_1^H/M$ &  0.33335970 &   0.33335041 &   0.09997350 &   0.09998022 &  0.04997869 &  0.04997759 \\
    $x_2/M$ & -9.95121484 &  -9.98358508 &  -10.3327592 &   -10.341799 &  -10.758263 &  -10.760757 \\
    $y_2/M$ &  3.29853493 &   3.29853493 &  -0.98956048 &   0.98956048 & -0.49478024 &  0.49478024 \\
  $p_2^x/M$ &  0.06465142 &   0.06465244 &   0.01947740 &   0.01948331 &  0.00973282 &  0.00973409 \\
  $p_2^y/M$ & -0.02082070 &   0.02264774 &  -0.02554774 &   0.02642299 & -0.01524037 &  0.01559326 \\
  $m_2^p/M$ &  0.32205500 &   0.32205500 &   0.09380000 &   0.09380000 &  0.04650000 &  0.04650000 \\
  $m_2^H/M$ &  0.33334372 &   0.33336827 &   0.09996202 &   0.10000287 &  0.04997149 &  0.04998916 \\
      $d/M$ &  6.59706986 &   6.59706986 &   1.97912096 &   1.97912096 &  0.98956048 &  0.98956048 \\
$d_{spd}/M$ & 9.19569237 & 9.19583305 &   2.88313183 &   2.88384645 &  1.45922435 &  1.45947139 \\
    $x_3/M$ &  19.9428396 &   19.9121871 &   2.59715370 &   2.58932285 &  1.20318845 &  1.20106818 \\
    $y_3/M$ &  0.00000000 &   0.00000000 &   0.00000000 &   0.00000000 &  0.00000000 &  0.00000000 \\
  $p_3^x/M$ & -0.00002303 &  -0.00002508 &  -0.00017086 &  -0.00018268 & -0.00007367 & -0.00007621 \\
  $p_3^y/M$ &  0.04300773 &  -0.04392914 &   0.05150537 &  -0.05243609 &  0.03068568 & -0.03098158 \\
  $m_3^p/M$ &  0.32911500 &   0.32906500 &   0.79320000 &   0.79320000 &  0.89620000 &  0.89620000 \\
  $m_3^H/M$ &  0.33336348 &   0.33334239 &   0.79961960 &   0.79964496 &  0.90002248 &  0.90002969 \\
  $M\Omega$ &  0.00582575 &   0.00586644 &   0.01925963 &   0.01938382 &  0.02139540 &  0.02147362 \\
      $D/M$ &  29.8940544 &  29.89577218 &   12.9299129 &   12.9311219 &  11.9614515 &  11.9618252 \\
\hline
\end{tabular}
\end{table*}

%% file: merger_times_massratio_table.tex
\begin{table}
\centering
\caption{Number of orbits to merger and merger time of the inner binary for different mass ratios with the third black hole. Merger time is normalised by the total mass of each inner binary $m$. Cases 3BH1-2q1-8-18}
\label{tab:3BHsq}
\begin{ruledtabular}
\begin{tabular}{lll}
Configuration & $\#$orbits & $t_{\text{merger}}/m$ \\
\hline
   3BH1q1  &      6.51 &            945.66 \\
   3BH2q1  &      6.62 &            972.28 \\
   3BH1q8  &      7.58 &           1461.25 \\
   3BH2q8  &      8.02 &           1688.75 \\
   3BH1q18 &      5.75 &           1595.00 \\
   3BH2q18 &      6.35 &           1785.00 \\
\end{tabular}
\end{ruledtabular}
\end{table}